\documentclass[sigconf]{acmart}

\AtBeginDocument{%
  }

\copyrightyear{2024}
\acmYear{}
\acmDOI{}

\acmConference[Pre-Print]{Arxiv}{September 16, 2024}{}
\acmISBN{}

\usepackage{tikz}

\begin{document}

\title{ArticulatePro: A Comparative Study on a Proactive and Non-Proactive Assistant in a Climate Data Exploration Task}

\author{Roderick Tabalba}
\email{tabalbar@hawaii.edu}
\orcid{0000-0002-6175-0353}
\affiliation{%
  \institution{University of Hawaii at Manoa}
  \city{Honolulu}
  \state{Hawaii}
  \country{USA}
}

\author{Christopher J. Lee}
\email{clee48@hawaii.edu}
\orcid{0009-0000-5180-5528}
\affiliation{%
  \institution{University of Hawaii at Manoa}
  \city{Honolulu}
  \state{Hawaii}
  \country{USA}
}

\author{Giorgio Tran}
\email{ttran2@hawaii.edu}
\orcid{0009-0001-9452-7828}
\affiliation{%
  \institution{University of Hawaii at Manoa}
  \city{Honolulu}
  \state{Hawaii}
  \country{USA}
}

\author{Nurit Kirshenbaum}
\email{nuritk@hawaii.edu}
\orcid{0000-0002-3587-1565}
\affiliation{%
  \institution{University of Hawaii at Manoa}
  \city{Honolulu}
  \state{Hawaii}
  \country{USA}
}

\author{Jason Leigh}
\email{leighj@hawaii.edu}
\orcid{0000-0001-7693-2814}
\affiliation{%
  \institution{University of Hawaii at Manoa}
  \city{Honolulu}
  \state{Hawaii}
  \country{USA}
}

\renewcommand{\shortauthors}{Roderick Tabalba, Christopher J. Lee, Giorgio Tran, Nurit Kirshenbaum, and Jason Leigh}

\begin{abstract}
Recent advances in Natural Language Interfaces (NLIs) and Large Language Models (LLMs) have transformed our approach to NLP tasks, shifting the focus towards a more Pragmatics-based approach. This shift enables more natural interactions between humans and voice assistants, which have been historically difficult to achieve. Pragmatics involves understanding how users often talk out of turn, interrupt one another, or provide relevant information without being explicitly asked (maxim of quantity). To explore this, we developed a digital assistant that continuously listens to conversations and proactively generates relevant visualizations during data exploration tasks. In a within-subject study, participants interacted with both proactive and non-proactive versions of a voice assistant while exploring the Hawaii Climate Data Portal (HCDP). Results suggest that the proactive assistant enhanced user engagement and facilitated quicker insights. Our study highlights the potential of Pragmatic, proactive AI in NLIs and identifies key challenges in its implementation, offering insights for future research.
\end{abstract}


\begin{teaserfigure}
  \includegraphics[width=\textwidth]{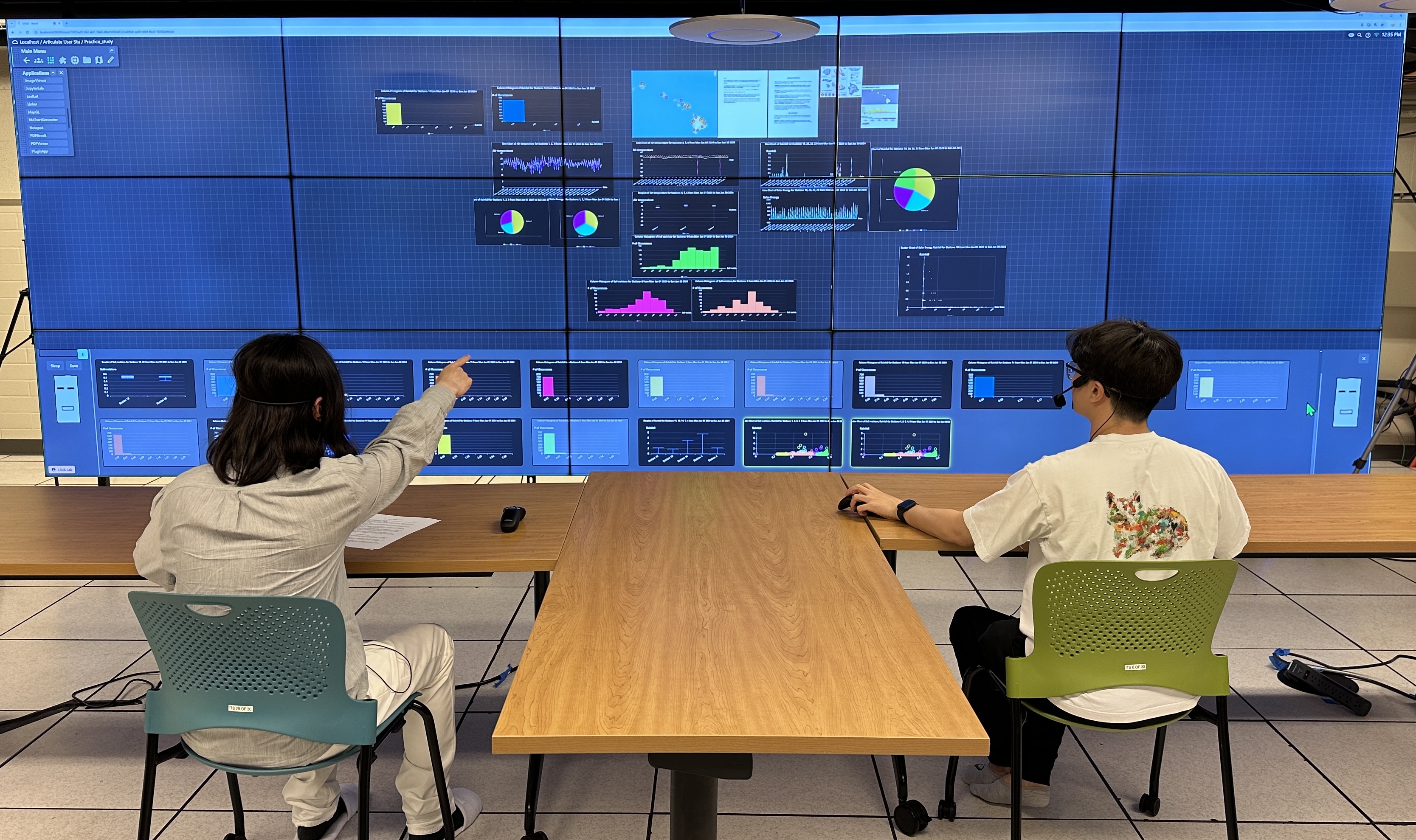}
  \caption{User Study setup where dyads of participants interacted with ArticulatePro, a visualization application integrated in the Smart Amplified Group Environment (SAGE3) software \cite{tabalba2023}. Participants talked to a voice-activated digital assistant to create visualizations on the Hawaii Climate Data Portal (HCDP) \cite{longman24, mclean2020hawaii, mclean2023building, mclean2023design}.}
  \Description{Two users sitting in front of a large tiled display wall with the software "ArticulatePro" open.}
  \label{fig:teaser}
\end{teaserfigure}

\received{16 September 2024}

\maketitle

\section{Introduction}
Advances in technology, such as Large Language Models (LLMs) like GPT \cite{achiam2023gpt}, BERT \cite{kenton2019bert}, and Llama \cite{touvron2023llama}, have enabled breakthroughs in various complex Natural Language Processing (NLP) applications. While researchers and policymakers are still addressing the ethical challenges posed by these powerful innovations, these general-purpose models have made it easier to tackle NLP tasks such as question answering, generating text in the style of deceased authors, and zero-shot or few-shot learning classification \cite{abbasiantaeb2024, pataranutaporn2023, bucher2024}. As a result, natural language research can now move beyond the focus on understanding semantics and syntax (as seen in text summarization and classification) and shift towards Pragmatics—understanding the nuances of natural language communication, such as text generation, natural language reasoning, and linguistic presuppositions \cite{li2021}. This shift gives us the opportunity to re-evaluate how we approach interacting with voice assistants to develop Pragmatics-focused Natural Language applications. For instance, a Pragmatic voice assistant could potentially understand the full range of how people naturally communicate, paving the way for innovations such as AI psychiatrists, elderly caretakers, and personal tutors.

In human conversations, proactivity involves natural interactions—interrupting, talking out of turn, and offering just the right amount of information \cite{balaraman2020proactive, semmens2019now}. Current home assistants, like Siri and Alexa, only respond to direct requests, and even advanced AI chatbots like ChatGPT act only when explicitly instructed. This results in AI mirroring the user’s thoughts rather than prompting new ideas or challenging their thinking, which can limit the depth of conversations \cite{wilson1986defining}. Prior research highlights the potential benefits and the growing demand for proactive AI \cite{volkel2021eliciting, meurisch2020exploring, van2021human}.

To explore the concept of a proactive AI assistant, we focused on the context of data exploration. Over the past decade, Natural Language Interfaces (NLIs) have made significant progress in assisting users with data exploration tasks \cite{shen2022}. These interfaces simplify the process of creating visualizations, enabling users to understand data without programming skills or expertise in visualization \cite{sun2010articulate, quadri2024}. Data exploration serves as our testbed for evaluating the effectiveness of our proactive digital assistant.

To address the gap in knowledge about proactive Artificial Intelligence, we pose the following research questions: 
\begin{enumerate} 
    \item What are the differences in user interaction and outcomes when using a proactive digital assistant versus a non-proactive digital assistant during data exploration tasks?  \item What are the benefits of interacting with a proactive digital assistant during data exploration? 
\end{enumerate}

To explore these questions, we developed ArticulatePro, a Pragmatic data visualization application featuring a digital assistant capable of proactively generating charts during data exploration tasks. The digital assistant continuously listens to conversations and detects opportunities to generate relevant charts at the right moment. We evaluated our approach through a within-subject user study, where dyads of participants interacted with both a proactive digital assistant and a non-proactive version of the assistant. By comparing participants' interactions in both conditions, our study aims to uncover the potential benefits and challenges of a proactive digital assistant. In our study, participants explored the Hawaii Climate Data Portal (HCDP), a climate data repository developed by the University of Hawaii at Manoa \cite{mclean2020hawaii, mclean2023building, mclean2023design, longman24}.

This paper makes the following contributions: 
\begin{enumerate} 
    \item A working prototype of a proactive system designed for data exploration tasks. 
    \item An evaluation of the effectiveness of our proactive digital assistant.
    \item A comparison of interactions between a proactive and a non-proactive digital assistant. 
    \item A discussion of the usability challenges encountered when working with a proactive digital assistant for data exploration tasks. 
    \end{enumerate}

\section{Relevant Works}
In this section, we describe the relevant research in natural language for data exploration applications, proactive digital assistants, and Pragmatic theories.

\subsection{Natural Language for Data Exploration}

Over the past decade, significant advancements have been made in the research of systems that facilitate data exploration tasks using natural language voice commands. This focus enables users to conduct data exploration more efficiently, allowing them to concentrate on the exploration process rather than on chart generation. In our prior work, we developed and evaluated Articulate \cite{sun2010articulate}, one of the earliest natural language-enabled data exploration systems \cite{shen2022}. We found that users, on average, created charts 12 times faster than when using Excel, showcasing the efficiency gains that can be achieved with a natural language system.

In 2015 and 2016, Gao et al. and Setlur et al. explored user interface elements that allowed users to manipulate their verbally-created charts \cite{gao2015datatone, setlur2016eviza}. These studies introduced hybrid interactions, where users could still utilize the keyboard and mouse alongside text interaction. However, this paper focuses primarily on voice commands for chart creation, rather than relying on other modalities.

In 2016, we conducted a Wizard-of-Oz study, revealing that only 15\% of user utterances were direct queries, while the remaining 85\% provided context for queries during data exploration tasks \cite{kumar2016towards}. Queries lacking sufficient context often failed to generate the intended visualizations. From 2018 to 2023, researchers including Hoque et al., Srinivasan et al., Setlur et al., and our own prior work, designed context-supported data exploration systems \cite{setlur2016eviza, srinivasan2021snowy, hoque2017applying, tabalba2022articulate, tabalba2023investigation}. These studies demonstrated how incorporating context could help repair imprecise or incomplete queries, emphasizing the importance of context-supported natural language systems—one of the key aspects of Pragmatics.

Between 2017 and 2023, we iterated on a conversational data exploration system, concentrating on resolving co-reference resolution \cite{aurisano2015show, aurisano2016articulate2, kumar2017towards, bhattacharya2023reference}. Large Language Models (LLMs) have been shown to perform exceptionally well in addressing co-reference resolution \cite{gan2024assessing}, showcasing how LLMs' generalized knowledge can be leveraged to solve previously challenging NLP problems.

\subsection{Proactive Applications}

The need for proactive systems has been widely discussed in the literature. In one study, McMillan et al. hired professional designers to analyze recorded daily conversations of users \cite{mcmillan2015repurposing}. The designers proposed potential applications that could utilize conversational data. One suggestion was to proactively launch mobile applications. For example, if keywords like "hungry" or "food" were detected by the mobile device, it could automatically launch Uber Eats, a popular food delivery application. Volkel et al. surveyed 205 participants on their vision of the "perfect voice assistant" \cite{volkel2021eliciting}. They found that the majority of participants wanted their voice assistant to take the lead in parts of the dialogue. Specifically, 87

Reicherts et al. conducted an online survey to explore user perceptions of storyboards featuring proactive voice assistants \cite{reicherts2021may}. Participants found the proactiveness of the assistant to be useful, appropriate, and pleasant. Similarly, Zargham et al. conducted an online study evaluating user perceptions of usefulness, appropriateness, and invasiveness \cite{zargham2022understanding}. They introduced the term "The Proactivity Dilemma" to describe the dual nature of Natural Language Interfaces (NLIs) initiating conversations—being both beneficial and potentially intrusive, depending on the context. This work highlights how opportunities for proactivity can affect users' perceptions of its intrusiveness. In our paper, we study how users interact with a working proactive assistant, rather than evaluating storyboards. Additionally, we focus on a specific instance of proactivity, comparing its effectiveness against a non-proactive assistant in a data exploration task.

Andolina et al. developed a system that could proactively retrieve Google search results, allowing users to focus on interpreting the results rather than formulating a query explicitly \cite{andolina2018investigating}. Our study differs by focusing on data exploration tasks with dyads of participants rather than retrieving Google search results with individual participants. Shi et al. developed a brainstorming system that could retrieve relevant images based on user conversations \cite{shi2017ideawall}. Their study demonstrated how a constantly listening system could help users generate more ideas and reduce the number of conversational lulls. While Shi et al. evaluated the effectiveness of displaying images, we compare how users interact with a proactive assistant versus a non-proactive assistant during data exploration tasks. Balaraman and Bernardo developed a proactive information retrieval system called SimDial \cite{balaraman2020proactive}, which, like our approach, focuses on Pragmatics but differs by evaluating short simulated dialogues. Their proactive approach reduced the number of back-and-forth exchanges between the user and the system. In our study, we implement a fully functional proactive digital assistant and evaluate its effectiveness in a real-world data exploration setting, as opposed to simulated dialogues.

\subsection{Pragmatics}

Li et al. provide a comprehensive overview of Natural Language Processing (NLP), highlighting three pivotal areas: semantics, syntax, and pragmatics \cite{li2021}. Their analysis indicates that while substantial progress has been made in semantics and syntax, there is an urgent need to shift NLP research toward Pragmatics, emphasizing its incorporation into the development of Natural Language Interfaces (NLIs).

Pragmatics, a branch of linguistics, deals with the nuances of natural language conversations. A key element of Pragmatics is context. According to Cutting, context encompasses the physical location of the conversation, the shared background knowledge, and the mutual understanding between participants \cite{cutting2005pragmatics}. The meaning of a sentence can change entirely depending on the context \cite{grice1975logic}. Paul Grice, a prominent figure in Pragmatics, introduced the concept of "implicatures," which extend beyond the semantic and syntactic dimensions of verbal or written communication. In our system, we incorporate context by continuously listening to the conversation, rather than only processing isolated user queries.

Grice also proposed the Cooperative Principles, which are guidelines that people typically follow during conversations. These principles include quantity, quality, relevance, and manner. The maxim of quantity suggests providing just enough information, while the maxim of quality emphasizes delivering information believed to be true. The maxim of relevance dictates that communication should be pertinent to the current discussion, and the maxim of manner advises that communication should be clear and unambiguous. It is important to note that these principles are based on interactions between cooperating individuals, and in real conversations, people may break these maxims for various reasons.

Sperber and Wilson further developed Grice's ideas, arguing that all conversations can be reduced to the maxim of relevance \cite{wilson1986defining}. They posit that this principle simplifies communication by focusing on what is relevant, which helps to establish common ground between conversing individuals. This common ground, referred to as the "cognitive environment," includes the shared knowledge between participants. In our system, we aim to apply Sperber and Wilson’s theory of relevance, ensuring that our system engages in meaningful and contextually appropriate interactions.

\section{ArticulatePro}
In this section, we describe an overview of how users interact with the application. Then we dive deeper into the technical components on how the system works internally.

\subsection{An Overview of the System}
ArticulatePro is implemented as an application in SAGE3 (Smart Amplified Group Environment), a web app designed for managing large tiled display walls \cite{tabalba2023, harden2023sage3}. SAGE3 is developed using JavaScript for the frontend and Node.js and Python for the backend. We chose SAGE3 as the foundation for ArticulatePro because it allowed us to leverage its features, such as managing content on an infinite canvas, while simplifying the development stack needed to build the application.

\begin{figure}[ht] \centering \includegraphics[width=\columnwidth]{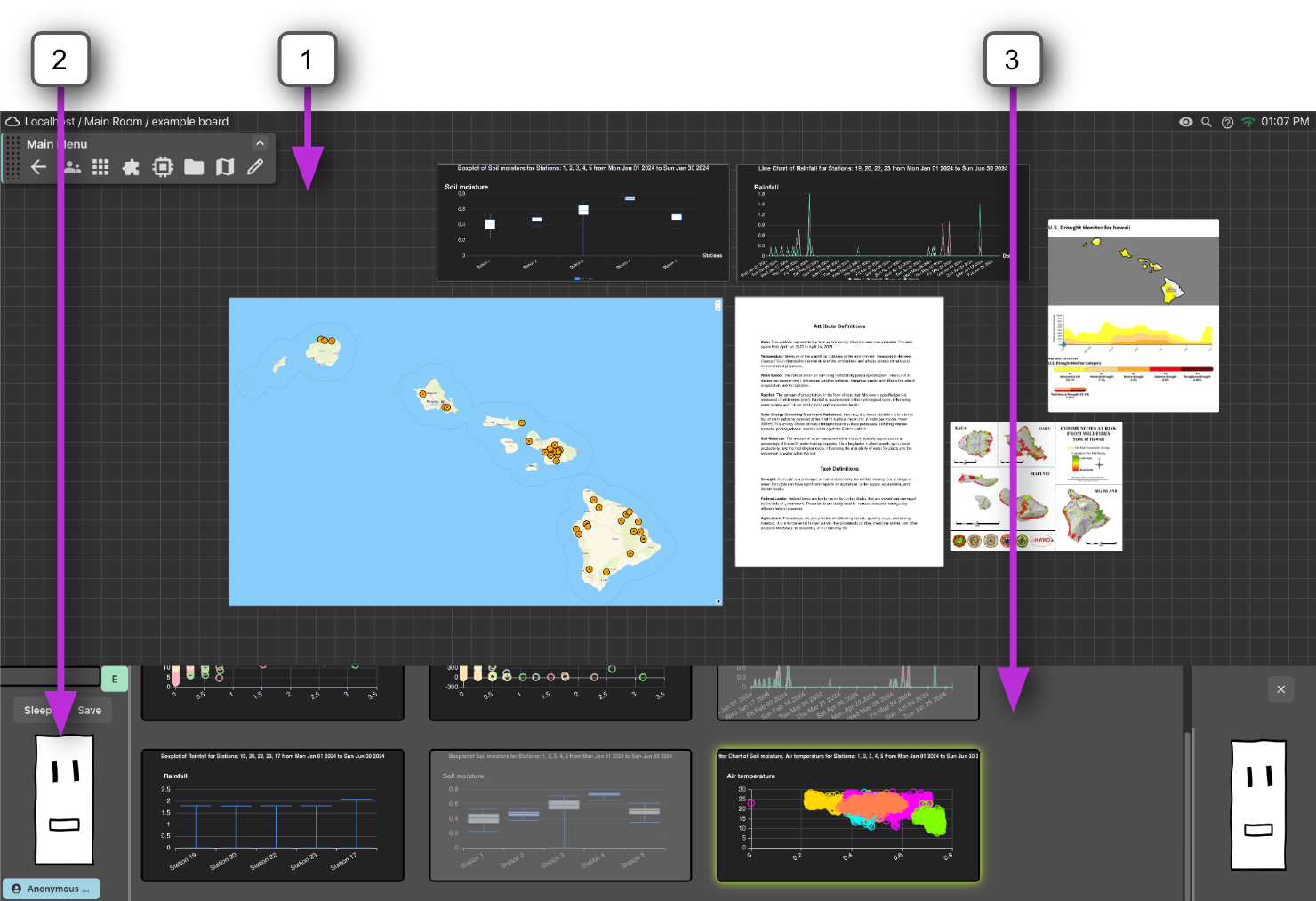} \caption{(1) General workspace for moving, resizing, and deleting selected visualizations. (2) The digital persona of ArticulatePro that creates visualizations. (3) A visualization conveyor belt (inspired by a sushi conveyor belt) that displays generated visualizations from the digital assistant. The assistant creates visualizations and displays them on the conveyor belt, where users can select charts to move them to the general workspace for further analysis.} \label{UI} \end{figure}

The ArticulatePro application, shown in Figure \ref{UI}, is a visualization tool where users can interact with a digital assistant, Arti, to assist in data exploration tasks using natural language voice commands. In our Pragmatic approach, we designed Arti to take advantage of conversational context by continuously listening to user conversations. Based on what Arti hears, it generates relevant charts. Context is a critical aspect of Pragmatics \cite{cutting2005pragmatics}, and by contextualizing user queries, Arti is able to resolve queries that may be incomplete or ambiguous. Arti determines the best time to generate visualizations, either in response to an explicit request or proactively, much like a human might respond in a conversation. Below are descriptions of an explicit request and how Arti proactively creates charts:

\begin{description} \item[Explicit Request] When the user explicitly asks for a chart, Arti behaves similarly to current digital assistants like Siri, Alexa, and Cortana, but without the need for wake words. Arti listens continuously and generates a chart whenever it detects a command. For example, the user might say, “Generate a chart on COVID risk and diabetes rate.” In this case, the user directly requests a chart that shows the two variables. \item[Proactive] In proactive instances, Arti generates a chart even when the user hasn’t explicitly requested one. For example, if two users are exploring medical data and one says, "I don’t get why COVID rates are higher in the East than in the West," Arti might generate a chart comparing medical factors between the East and West, even though the user didn’t directly ask for it. This could help clarify the user’s confusion by providing additional context on the discrepancy in COVID rates between the East and West. This example illustrates how Arti can be proactive during data exploration tasks. \end{description}

Once Arti generates a chart—either through an explicit request or proactively—it is displayed on the visualization conveyor belt (Figure \ref{UI}, Box 3). Users can hover over the chart to see a larger preview in the center of the screen for easier inspection. If they find the chart valuable, they can click on it to move it to the workspace (Figure \ref{UI}, Box 1), where it can be rearranged, resized, compared with other charts, or deleted.

In the following sections, we describe how we designed our proactive approach. First, we explain our investigation into exploring opportunities for a digital assistant to be proactive during data exploration tasks. Then, we describe our proactive solution, leveraging Pragmatic theories of communication.

\subsubsection{Investigating Proactive Opportunities}

To explore the various opportunities for a digital assistant to be proactive, we conducted an analysis based on our prior work \cite{tabalba2023investigation}. In that study, we investigated the benefits of an always-listening digital assistant by having dyads of participants interact with it during data exploration tasks. This study allowed us to observe potential moments where a digital assistant could be proactive. This step was essential for our research, as no prior work in the literature has focused on investigating proactive opportunities in data exploration tasks. For our analysis, one researcher reviewed the study's utterances and developed classifications for situations where a digital assistant could potentially be proactive. The following classifications were identified:

\begin{description} 
    \item [Discovery] Users often verbally state their findings or discoveries. For example, a user might examine a chart and say, "So as COVID rates increase, so does poverty," or "Social vulnerability seems to overlap with the map of poverty." This presents an opportunity for the assistant to confirm or challenge the user’s discovery. 
    \item [Disagreement] Users might express disagreement with one another, saying things like, "I don’t think that’s true," or "The chart doesn’t necessarily show that." In such cases, the assistant could provide relevant data to resolve the disagreement. 
    \item [Preference] Users may state their preference for the type of chart they want to see, such as, "I think a good strategy is to use a lot of maps." Here, the assistant could remember the user's preference and prioritize similar chart types in the future. 
    \item [Criticism] Users may criticize the charts provided to them, saying things like, "What does that say? I'm blind, I can't read the text." In such cases, the assistant could adjust the chart by increasing font size or making other enhancements. 
    \item [Curiosity] Users may indicate what they plan to explore next, saying something like, "It seems like the Southeast and small cities would be good places to start. Let’s prioritize resources based on these variables." This is an opportunity for the assistant to generate relevant charts based on the user’s interests before they explicitly ask for it.
    \item [Confusion] Users might express confusion about the task at hand, saying something like, "When we talk about prioritizing resources, what are resources? Do doctors count as a resource?" In such moments, the assistant could step in to clarify or offer guidance. 
\end{description}

These classifications are consistent with the proactive assistant storyboards envisioned by Zargham et al. and Reicherts et al. \cite{reicherts2021may, zargham2022understanding}. However, while their classifications were based on home assistants, ours are focused on the context of data exploration.

Out of the identified proactive classifications, we decided to focus on the most prevalent one: when users make a finding or discovery. This classification was the most common in our study and, by selecting it, we maximized the chances of observing user reactions to a proactive assistant while limiting the complexity of our study to a single aspect of proactivity.

\subsubsection{Pragmatic Approach for Proactive Opportunities}

To guide the development of our proactive assistant, we applied Pragmatic theories. According to Grice’s principles of Cooperative Conversation \cite{grice1975logic}, communication is most effective when it adheres to the principle of relevance \cite{wilson1986defining}. Based on this, we incorporated a "history" component into our assistant, enabling it to track the conversation, the user’s interactions, and the charts it has already generated. This allows the system to produce relevant charts based on the ongoing dialogue and user preferences while minimizing redundancy. For more details on our implementation of the history component, see Section \ref{history}.



\subsection{Technical Overview}

\begin{figure}[ht]
  \centering
  \includegraphics[width=\columnwidth]{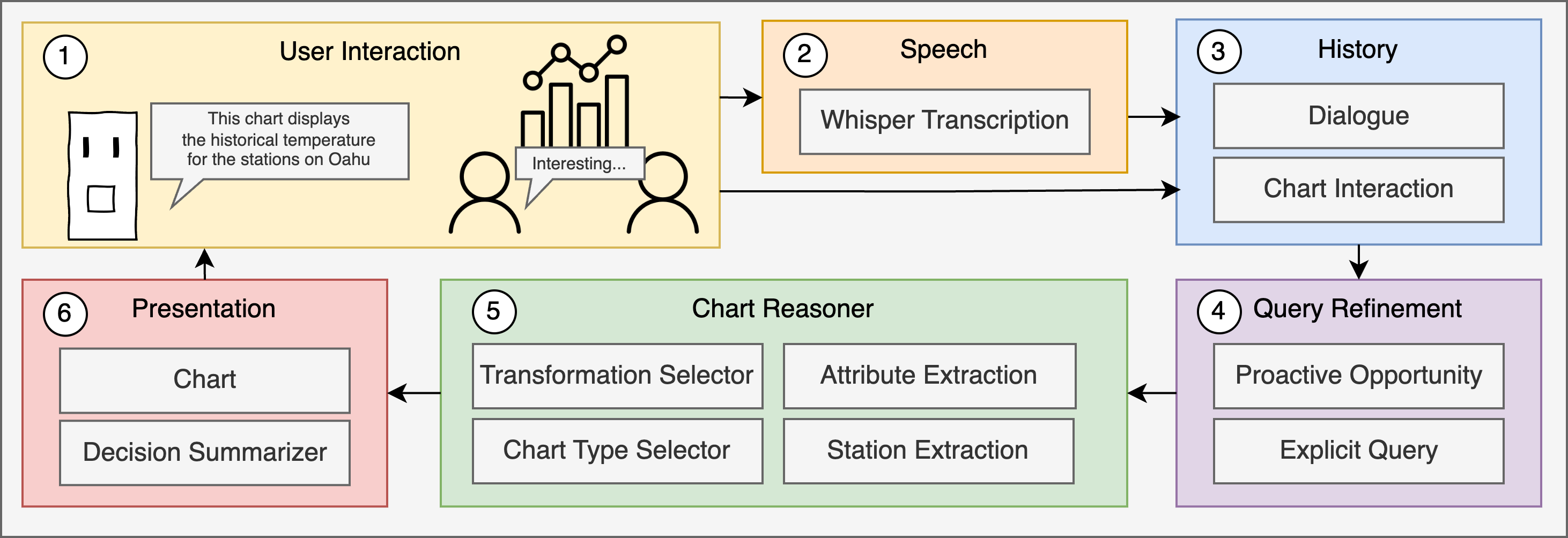}
  \caption{The architectural design of ArticulatePro. (1) User interaction consists of the user's speech and interaction of charts. (2) Speech module transcribes the user's speech using OpenAI's Whisper model \cite{radford2023robust}. (3) History consists of a history of the user's utterance and chart interaction. (4) Query Refinement creates a succinct query based on the history component. (5) Chart Reasoner extracts and decides what chart to construct and what attributes to visualize (6) Presentation component for chart construction and response generator.}\label{diagram}
\end{figure}

In this section, we describe the internal components on the ArticulatePro application. Figure \ref{diagram} displays the architecture of ArticulatePro. 

In the following sections, we describe each of the numbered boxes shown in the Figure. As the arrows indicate, each box represents a step in the pipeline, passing the output from one box as the input to the next box. The pipeline starts with the user’s query and chart interaction.

\subsubsection{User Interaction}
This is where the user interacts with the digital assistant. In this module, the assistant presents charts to the user along with textual content summarizing what it has generated. The assistant tracks what the user says, which charts they select, and how they interact with them. When the system detects a noise, it sends the audio to the speech component for processing.

\subsubsection{Speech}

In our prior work, we found the Web Speech API produced many incomplete utterances, likely due to the API's parameters duration of silence to detect the end of an utterance \cite{tabalba2023investigation}. While the Web Speech API may work better for natural language applications that are not continuously listening to the user, we required a speech component that could support an always-listening environment. Therefore we designed a custom speech component using the OpenAI's Whisper model \cite{radford2023robust}. 

From our prior work of when people talked to an always-listening digital assistant, we calculated the average pause lengths between user utterances. On average, we found that it took users on 1.4 seconds to complete their utterance. To improve recognition on detecting complete utterances, we increased the pause length for our speech component to 1.5 seconds. Therefore, once the system detects noise followed by 1.5 seconds of silence, it will attempt to transcribe the audio using the Whisper base English model from OpenAI. If the audio is recognized as text, it is stored in the History component for later use.

\subsubsection{History} \label{history}

ArticulatePro maintains two types of histories:

\begin{enumerate}
    \item Dialogue History
    \item User Interactions, which include: 
    \begin{itemize}
        \item Charts that the user selected
        \item Charts that the assistant generated
        \item The last chart that the user interacted with
    \end{itemize}
\end{enumerate}

According to Pragmatics, keeping track of the context is crucial for building a successful pragmatic natural language application. Here are specific implementation details on how the system keeps track of history:

\begin{description}
    \item[Dialogue History] ArticulatePro limits the dialogue history to the last five utterances. We experimented with longer context lengths, but this led to the model overemphasizing irrelevant context. For example, early in a data exploration task, the user might say, "We need to focus on rainfall for renewable energy." Later, they might decide that wind speed and solar energy are more important. If the model retains the initial focus on rainfall, it could generate less relevant content. In our prior work, we found that a five-utterance history strikes a good balance between relevance and filtering out outdated content \cite{tabalba2023investigation, tabalba2022articulate}. While other dialogue management strategies could be used, this approach meets our needs.
    \item[User Interaction History] The user interaction history keeps track of the charts that the user interacts with and what the digital assistant generated. This history provides context that allows the system to generate relevant charts and avoids generating duplicate charts. We keep track of the charts by storing the titles of the charts. We limit the history to the most recent 5 charts the user selected, most recent 5 charts that the assistant generates, and the last chart that the user selected. 
\end{description}

This history is then utilized in the query refinement component.

\subsubsection{Query Refinement} \label{refinement}

In the query refinement component, ArticulatePro rewrites the user’s query in the context of the history. This allows the system to make charts that are relevant, correct incomplete or inaccurate utterances, enhance accurate, and minimize the chance of producing redundant visualizations.

ArticulatePro detects whether the user has made an “Explicit Query” or a “Proactive Opportunity” using a neural network we trained. Details on our neural network training can be found in the next Section \ref{neuralNetwork}.
Here are descriptions on what the system does when it detects "Explicit Query" or "Proactive Opportunity" 
\begin{description}
    \item[Explicit Query] Explicit queries are utterances where the user directly asks for a chart. Here are some examples:
    \begin{itemize}
        \item Generate a chart on the solar energy for the Big Island.
        \item Show us a graph of the air temperature on Oahu.
        \item Display a chart of the highest recorded rainfall measurement in Hawaii.
    \end{itemize}
    \item[Proactive Opportunity] Using a classification model that we trained, ArticulatePro continously listens to the conversation and classifies each user utterance. If the system classifies an utterance as a discovery/finding, it will try to proactively generate a chart. Once a proactive opportunity is detected, the system creates a new query based on the history. Here are a few examples of findings or discoveries that the system would detect:
    \begin{itemize}
        \item Station 4 on Oahu has the most rainfall.
        \item So as fuel efficiency increases, so do sales.
        \item There are a lot of affordable properties in rural areas.
    \end{itemize}
    \item [Non-Query] If a query is not detected as either explicit or proactive, it is labeled as a non-query. ArticulatePro does not perform an action on non-queries.
\end{description}

If the user's utterance is classified as either "Explicit Query" or "Proactive Opportunity", the LLM is told to rewrite the user's query to produce a useful and relevant visualization based on what is included in the history.

\subsubsection{Neural Network Training} \label{neuralNetwork}

ArticulatePro detects these types of queries using a neural network that we trained. For our training data, we used utterances collected from our prior study of users interacting with an always-listening data exploration assistant. One researcher manually labeled the utterances as either “Explicit Query,” “Proactive Opportunity,” or “Non-Query.” However, we had a limited number of training utterances from the study, so once we identified common utterance types, we used GPT-4.0 to generate additional training examples. This is a common approach for generating training data for NLP tasks \cite{schmidhuber2024llm}.

We trained the neural network classifier to predict the class labels from input embeddings. The model architecture consists of two fully connected layers:
\begin{description}
    \item[Input Layer] A fully connected layer with an input size of n, where n is the dimensionality of the input embeddings.
    \item[Hidden Layer] A fully connected layer with 128 hidden unites followed by a ReLU activation function.
    \item[Output Layer] A fully connected layer with 3 output units: Proactive Opportunity, Explicit Query, and Non-Query
\end{description}
We used the Adam optimizer with a learning rate of 0.001 and CrossEntropyLoss as the loss function. The model was trained for 20 epochs, with the dataset split as follows:
\begin{description}
    \item[Training Set] 60\% of the dataset
    \item[Validation Set] 20\% of the dataset
    \item[Test Set] 20\% of the dataset
\end{description}
The classifier was trained using a batch size of 32. At each epoch, the training loss was calculated by averaging the CrossEntropyLoss over all batches in the training set. The validation loss was similarly calculated by averaging the loss over all batches in the validation set. The model's weights were saved at the end of each epoch to allow for potential model selection based on validation performance. At Epoch 6, the training loss was 0.676, and the validation loss was 0.670, indicating that the model was not significantly overfitting and that its performance on the validation set was comparable to the training set. Finally, we tested the model on the test set and achieved 93.3\% accuracy.

Once a query is detected as either an explicit or proactive opportunity, ArticulatePro rewrites the query based on the history. The rewritten query is then sent to the Chart Reasoner component.

\subsubsection{Chart Reasoner}

The Chart Reasoner generates the details needed to create the user's chart. Lee et al. introduced the term "Macro-Query," defined as a broad or high-level request to a computer that lacks explicit instructions on what data attributes to retrieve or what transformations to apply \cite{lee2024macro}. These queries move beyond simple syntactic retrieval and involve semantic reasoning and implicit information retrieval. In our work, we address this notion of "Macro-Query" by developing a series of large language models (LLMs) designed to reason through tasks. The system then uses this reasoned information to assist in chart construction.

Here is a list of tasks the system addresses:

\begin{description}
    \item[Attribute Extractor] The Attribute Extractor identifies and extracts the necessary attributes from the user’s query.
    \item[Station Extractor] ArticulatePro retrieves data from the Hawaii Climate Data Portal, which contains climate data from stations across Hawaii. In this step, the system filters specific stations that fit the user’s query.
    \item[Transformation Selector] The Transformation Selector applies the necessary filters to the data based on the user’s query.
    \item[Transformation Selector] Based on what attributes the system has extracted from the query, the chart type selector will choose the best chart to represent the user’s query.
\end{description}
The Attribute Extractor, Station Extractor, and Transformation Selector follow no particular order. The only task that depends on the prior steps is the Chart Type Selector. This is because some chart types require certain attributes. For example, a scatter plot requires 2 nominal data attributes in order to be created. 

For each extraction and selection component, we use an LLM to complete the task. The models chosen for each task were selected based on two factors: speed and accuracy. We experimented with the following LLMs: gpt-4o-2024-05-13, Llama3-70b-8192, Llama3-8b-8192, mixtral-8x7b-32768, Gemma-7b-it, and gpt-3-turbo. Based on our experiments, we decided to use a combination of Llama3-70b and GPT-4o-2024-05-13. Llama3-70b-8192 performed the tasks fastest; however, it occasionally struggled to generate consistently accurate and reliable answers. For tasks where Llama3-70b was inconsistent, we used the next fastest model that could deliver reliable results, GPT-4o-2024-05-13. Specifically, we chose to use Llama3-70b-8192 for the Attribute Extractor and Chart Type Selector and GPT-4o-2024-05-13 for the Station Extrator and Transformation Selector.

We also employed a prompt engineering technique that includes adding a reasoning step, as described in the paper \cite{sahoo2024systematic}. The reasoning step was added for two reasons:
\begin{enumerate}
    \item Performance Improvement: Prior research has shown that asking the LLM to justify its answers can increase the system's performance.
    \item User Feedback: At the end of the decision-making process, the reasoning steps, along with the decisions, were summarized by a specialized agent. This summary was then presented to the user as 1 or 2 sentences.
\end{enumerate}
\subsubsection{Presentation}
In the final component, the system produces two outputs: the chart for the user and a summary of what it has generated. ArticulatePro uses the information gathered from the Chart Reasoner component to construct the chart.

First, the system uses the stations extracted by the Station Extractor and the attributes from the Attribute Extractor to fetch the necessary data. Next, it applies the transformations selected by the Transformation Selector. Finally, based on the chart type chosen by the Chart Type Selector, the system utilizes pre-defined code to construct the chart and fills in the required details. For this process, we use the ECharts library by Apache \cite{li2018echarts}.

This final component also generates a summary of its decisions for the user. It uses the information gathered from the reasoning step in the Chart Reasoner component. We use the LLM Llama3-70b-8192 to summarize these decisions.

\section{Evaluation}
In this section, we outline our evaluation methods to address the following research questions:
\begin{enumerate}
    \item What are the differences in user interaction and outcomes when using a proactive digital assistant versus a non-proactive digital assistant during data exploration tasks?
    \item What are the benefits of interacting with a proactive digital assistant during data exploration?
\end{enumerate}

To answer these questions, we conducted a comparative within-subject user study where participants interacted with two conditions: P where the proactive version of ArticulatePro was used and NP, where the non-proactive version of ArticulatePro’s digital assistant was used. To simplify these conditions to the participants, the agent in the P condition was referred to as “Arti”, and the agent in the NP condition was referred to as “Marti”. Using these names helped the participants differentiate between the two conditions as well as added to the anthropomorphization of the agents. 

\subsection{Answering Research Question 1}

To evaluate interaction for Research Question 1, we defined interaction metrics as:
\begin{enumerate}
    \item Total number of utterances
    \item Total number of task-relevant keywords spoken
\end{enumerate}

To evaluate outcomes for research question 1, we defined an outcome metric as the total number of “good utterances” that occurred during the session. We defined the meaning of a “good utterance” using a codebook developed with three researchers from a data visualization laboratory. We defined a "good utterance" as reflecting a discovery, insight, finding, or decision related to the dataset. See Section \ref{analysis} (Discovery Analysis) for more details.

\subsection{Answering Research Question 2}

To evaluate Research Question 2, we designed a semi-structured post-interview for the participants. The interview allowed us to gain insight on what benefits the proactive assistant had. In our discussion, we interpreted the participants' feedback in conjunction with the metrics from Research Question 1 and our own observations. This allowed us to verify the insights that we gathered with quantitative evidence that we measured. 

\subsection{Participants}
We recruited 24 participants grouped into dyads, forming 12 groups. Participants were aged 20 - 47, with backgrounds in Computer Science, healthcare, and engineering. This study was approved by an ethics board and all participants consented to participate in the study and be audio and video recorded. Participants were compensated with a /$30$ Amazon gift card.

\subsection{User Study Tasks}
For this study we used a subset of the data provided by the HCDP. The HCDP \footnote{https://www.hawaii.edu/climate-data-portal/} provides high-quality reliable climate data, such as temperature and rainfall, for the state of Hawaii. This data is gathered from a large network of climate sensor stations deployed around the state's islands. The subset of the data we used limited the data available to participants and included only 6 months of data from January 1st, 2024 to June 31st, 2024, taken from only a sampling of the sensor stations available in every island: 3 from Kauai, 3 from Oahu, 1 from Molokai, 10 from Maui, and 16 from the Hawaii, totaling in 33 stations that were used during this study. The data attributes that we made available to participants included rainfall, temperature, soil moisture, solar energy (incoming shortwave radiation), and wind speed. In addition to this queryable data, the SAGE3 board presented to participants was preloaded with images depicting drought and historical fire risk data around the state of Hawaii. These images were added to simulate a more realistic scenario.

We designed two tasks to be performed using the provided dataset. Here is a short description of the tasks:
\begin{enumerate}
    \item Imagine you are a farmer in Hawaii looking for good agricultural land to grow crops and raise cattle. Based on what charts the digital assistant generates, identify areas that you would consider for good agricultural land to grow crops and raise cattle.
    \item Imagine you are responsible for planning land usage in Hawaii for renewable energy. Based on the charts that the digital assistant generates, identify areas in Hawaii that may be good for wind farms and solar panel energy.
\end{enumerate}

Each task was allotted 30 minutes to complete. Participants were asked to prepare an answer for each task with supporting visualizations. To counterbalance the study, the order of tasks and conditions were alternated. After every 2 sessions, the order of the tasks were switched. For example, groups 1 and 2, the participants started with task 1 and ended with task 2. For the next 2 sessions, 3 and 4, the participants started with task 2 and ended with task 1. We continued to switch the order of tasks every 2 sessions. For the order of the conditions, the odd numbered groups were tested with condition P (Arti) followed by condition NP (Marti), and the even numbered groups tested with NP first, followed by P. By switching the ordering of the tasks and conditions this way, we were able to evaluate whether the ordering of conditions or tasks affected the results of the study.

\subsection{User Study Procedure}

In this section, we describe the user study procedure.

Participants in the study were provided with the following setup (one for each participant):
\begin{itemize}
    \item Two mice
    \item Two desks (72 x 30 inches)
    \item Two chairs
    \item Two Shure headset microphones
    \item Two pens
    \item Two copies of the task instructions
    \item Two copies of the attribute definitions
\end{itemize}

Participants shared a large tiled display wall (81 x 205 inches) as their focal point and substrate for performing the study. Both mice were connected to this display wall, however, only one would be active at any given time (single cursor), a limitation which may have affected some participants' workflow in regards to mouse-based interactions.

The User Interface created for ArticulatePro within SAGE3 (See Figure 1), includes a panel with a Visualization Conveyor belt where the system would render charts according to its active agent's behavior. The participants can hover over each visualization to see an enlarged preview, or click on the visualization to open it as an app within the SAGE3 board.

Participants were seated at desks approximately 30 inches apart to encourage loud enough communication for the digital assistant and the researcher to hear. They were fitted with headset microphones before the study began. The researcher explained about the overall structure of the study. They then introduced participants to the agent that corresponds to the assigned condition and the task they are asked to perform in the subsequent 30 minutes. The researcher reminded participants to discuss their observations with both each other and the digital assistant. After the 30 minutes elapsed, the participants were asked to provide a solution for the task, and back it up with visualizations. The remaining condition and task followed a similar procedure.

Before participants worked on any of the tasks, they were guided through a training session.

In this session, participants were shown examples of the chart types the system can generate (bar chart, line chart, specify the rest) illustrated with data from a known car information dataset \cite{srinivasan2021collecting}. The participants then practiced interacting with the system and SAGE3 board, performing actions such as:
\begin{itemize}
    \item Moving an application by clicking and dragging its window
    \item Resizing an application by adjusting its corners
    \item Panning the board by clicking and dragging the background
    \item Zooming in and out using the scroll wheel
    \item Selecting an app to interact with it
    \item Deleting an app
\end{itemize}

We deliberately avoided training participants on how to ask for charts to capture their natural interactions, as Srinivasan et al. \cite{srinivasan2021collecting} found that providing example queries can limit users to those examples.

\subsection{Participant Feedback}

After the completion of both tasks, participants were asked to individually fill-out a brief questionnaire assessing their attitude towards Arti and Marti. This questionnaire asked about their personal experience with each of the digital assistants; They were asked to rate on a scale from 1 to 5 whether the assistant was annoying when it interjected with charts, whether the assistant produced relevant content, and whether the assistant produced useful content. They were also asked to choose their preferred agent, Arti or Marti.

This was followed by a joint oral semi-structured interview. Participants were asked for their demographics (including age, major, and experience with visualization) and were guided in providing qualitative feedback based on their individual rates and preferences.

\section{Results}
This section presents the quantitative and qualitative results from the study and their method of collection.

\subsection{Total Number of Utterances}
We measured the number of utterances during each session, defined by our speech recognition module. When noise is detected, the speech recognition starts recording. After a 1.5-second pause, the recorded audio is transcribed. If the transcription results in written text, it is counted as an utterance.

\begin{figure}[ht]
  \centering
  \includegraphics[width=\columnwidth]{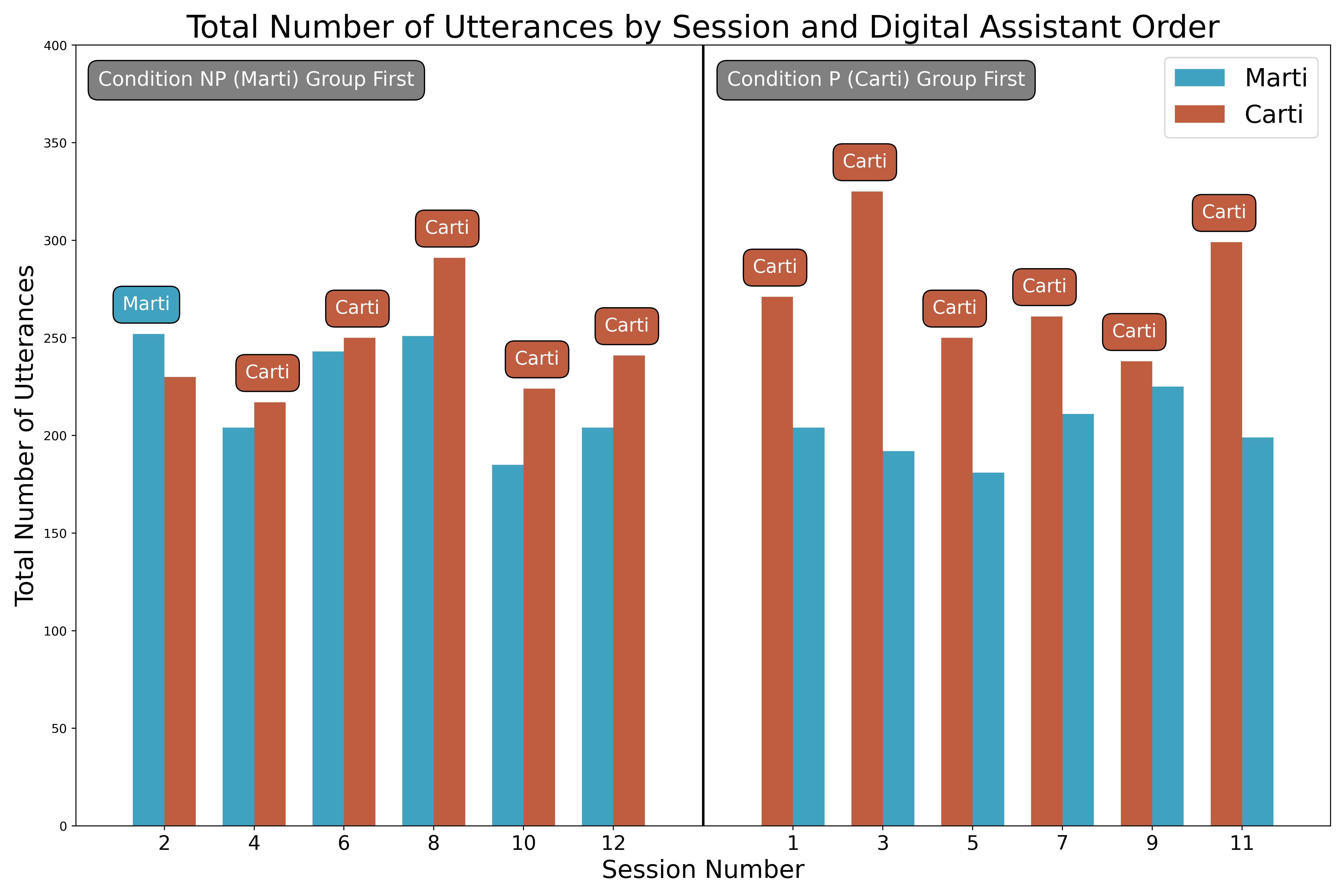}
  \caption{This chart displays the total number of utterances for each group. The chart is separated between the two groups P (Arti) first and NP (Marti) first. In almost every session but 2, users talked more with Arti.}
  \label{utterances}
\end{figure}

Results of a right-tailed paired-t test indicated that there is a significantly large difference between number of utterances towards Marti (M = 212.6 ,SD = 24.7) in the NP condition and number of utterances towards Arti (M = 258.1 ,SD = 32.9) in the P condition, t(11) = 3.7, p = .002*. However, graphing our results showed us (see Figure \ref{utterances}) the potential for a significant order effect despite our balancing of conditions.
With this discrepancy in mind, we further analysed the data by separating the participating groups into the groups who tested with Marti first followed by Arti (NP->P) and groups who tested with Arti first followed by Marti (P->NP). A right-tailed paired-t test for the (NP->P) group showed no significant results between utterances towards Marti (M = 223.2 ,SD = 29) and utterances towards Arti (M = 242.2 ,SD = 26.7), t(5) = 1.9, p = .059, while the right-tailed paired-t test for the (P->NP) group indicated that there is a significant large difference between utterances towards Marti (M = 202 ,SD = 15.3) and utterances towards Arti (M = 274 ,SD = 32.5), t(5) = 4.3, p = .008*. Further analysis with Mixed Repeated two-way ANOVA, confirms no significant effect between the groups (P->NP) and (NP->P), a significant effect within subjects F(1,10)=21.5464, p<0.0001**, and a significant interaction effect between the groups and number of utterances with Marti and Arti F(1,10)=7.3088, p=0.022*.

\subsection{Keyword Analysis}

The keywords chosen for the keyword analysis include the dataset's variable names: temperature, wind, rainfall, solar, and soil. We also included task-related keywords such as station, fire, drought, farm, and agriculture. Based on our internal discussions in our visualization laboratory, these keywords were deemed most relevant to the task. By performing a keyword analysis, we can determine whether the additional recorded utterances were relevant to the task.

\begin{figure}[ht]
  \centering
  \includegraphics[width=\columnwidth]{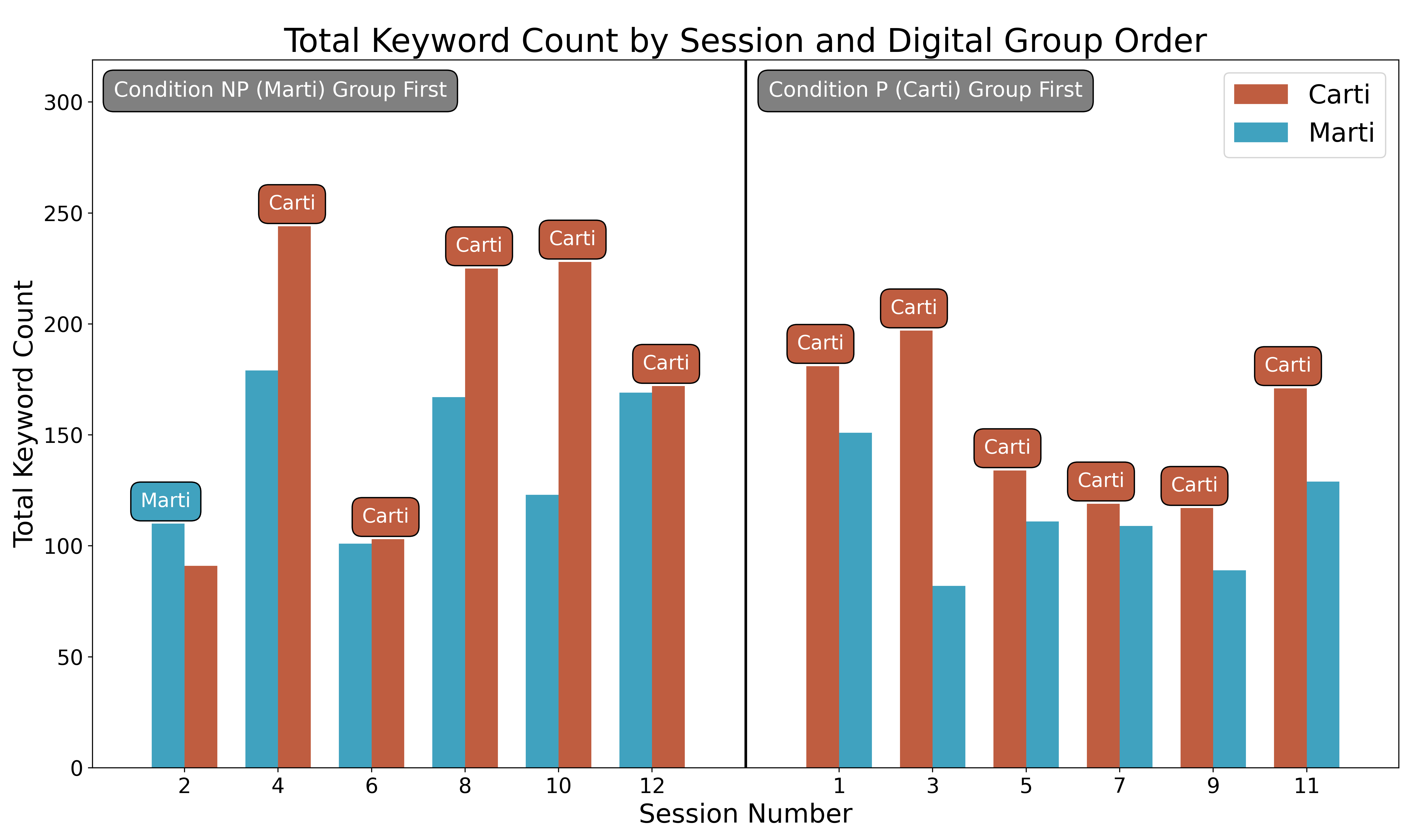}
  \caption{This chart displays the total number of utterances for each group. The chart is separated between the two groups P (Arti) first and NP (Marti) first. Similarly in the total utterance section, users used more task-relevant keywords in almost every session but session 2.}
  \label{keywords}
\end{figure}

Figure \ref{keywords} shows the total number of keywords mentioned during the tasks. As seen in the figure, in every session except session 2, users used more keywords with Arti than with Marti. Using Mixed Repeated two-way ANOVA showed no significant effect to the order of conditions nor an interaction effect in contrast to the total number of utterances, however, there was a significant effect in the number of keywords mentioned to Marti (M = 126.6, SD = 31.19) and the number of keywords mentioned to Arti (M = 165.16, SD = 49.89) with F(1,10) = 9.6633, p = 0.0110*.

\subsection{Discovery Analysis} \label{analysis}

 In this section, we aim to measure the outcomes of the data exploration task. We defined outcomes as the number of “good utterances” that occurred during the session. Defining a "good utterance" is subjective, so we developed a code-book with three researchers from a data visualization laboratory to establish criteria for this. To create the code-book, we selected a random subset of utterances from all sessions. The subset size was calculated using a sample size calculator for a 95\% confidence level with a 5\% margin of error, resulting in 355 utterances. Each researcher individually reviewed these 355 utterances to define what constitutes a "good" utterance.
 
The three researchers then met to discuss commonalities in their definitions and agreed on the following criteria for a "good" utterance:
"The utterance reflects a discovery, insight, finding, or decision related to the dataset. For example, a user may indicate that a station, area in Hawaii, or an entity (using a pronoun like "that" or "it" is sufficient, lacking, or comparable on any of the task-related variables (rainfall, temperature, soil moisture, etc..). This criterion is based on the likelihood that these utterances resulted from users examining a chart generated by the digital assistant." For simplification, we will refer to a "good" utterances as a "discovery".

Here are examples of discoveries that the participants made:
\begin{itemize}
    \item I'm just gonna set solar energy, so 18. Well, 18 for highest solar, yeah..
    \item The foundation 20 has a better average than 14 does. Cause ithen if you go--Oh yeah, there's some lows.
    \item I feel like station 20 should be a good pick. Right? But it has lower soil moisture
    \item The variability is lower in 23
    \item But 21 has highest solar and wind speed. I think we saw that somewhere.
\end{itemize}

The utterance must contain enough contextual information to be understood. Here are some examples of utterances lacking sufficient context:

\begin{itemize}
    \item So it could be used for..
    \item Rainfall for Maui stations.
    \item There's only one site.
\end{itemize}

We also ignored possible inaccuracies from the speech recognition system where feasible. 
Once the final codebook was established, one researcher manually labeled all 5,500 utterances. To minimize bias during the labeling process, the utterances were randomized and the researcher only had access to the utterances. 

\begin{figure}[ht]
  \centering
  \includegraphics[width=\columnwidth]{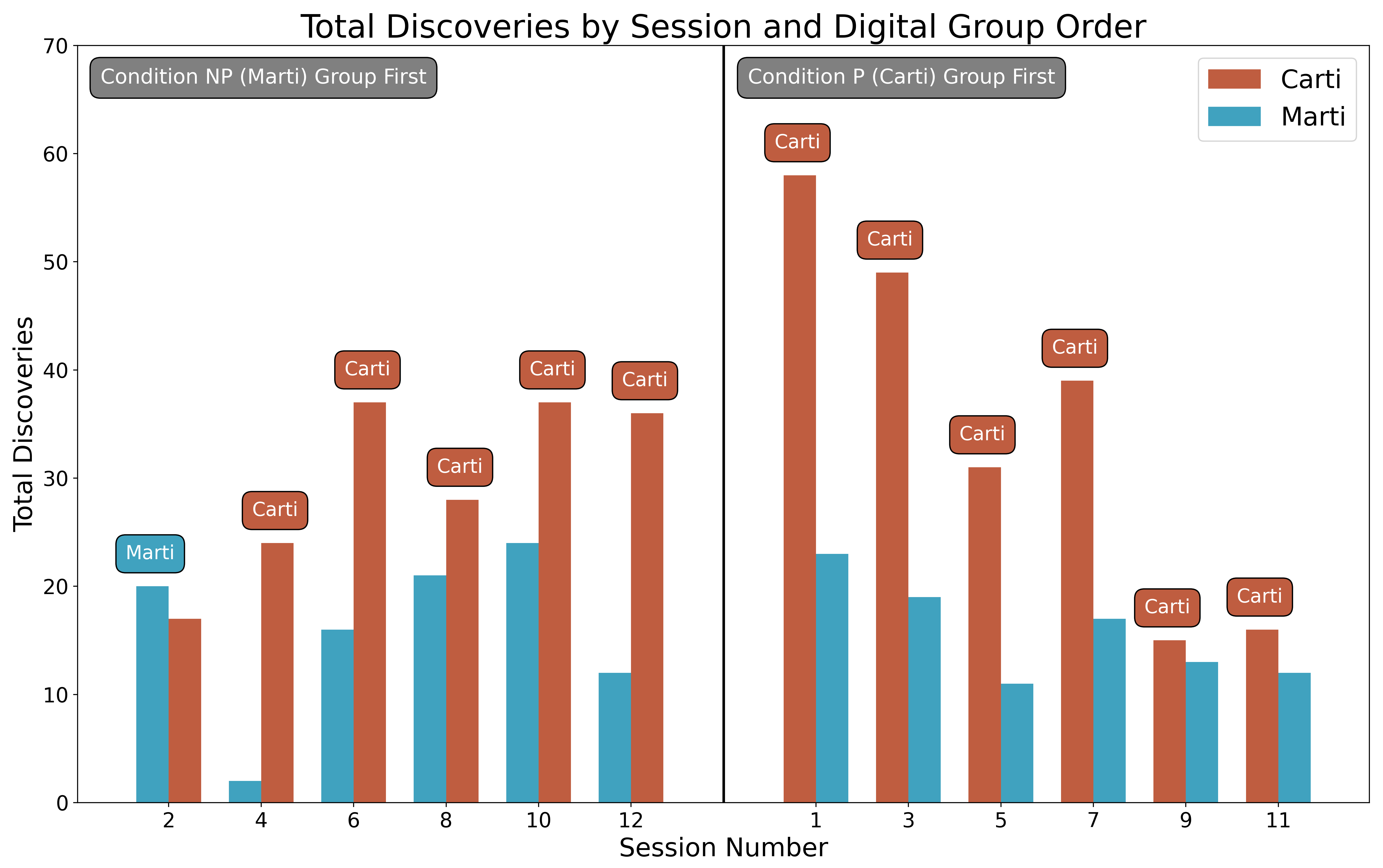}
  \caption{This chart displays the total number of discoveries for each group. The chart is separated between the two groups P (Arti) first and NP (Marti) first. In almost all sessions but session 2, the participants made more discoveries with Arti than with Marti.}
  \label{discoveries}
\end{figure}

Figure \ref{discoveries} shows the total number of "good utterances". As seen in the figure, in every session except session 2, users used more keywords with Arti than with Marti.

From session 1 to 12, the total number of discoveries that occurred in the sessions with Arti were: 58, 17, 49, 24, 31, 37, 39, 28, 15, 37, 16, 36 (\textit{M} = 32.25, \textit{SD} = 13.25), respectively. The total number of discoveries that occurred in the sessions 1 to 12 with Marti were: 23, 20, 19, 2, 11, 16, 17, 21, 13, 24, 12, 12 (\textit{M} = 15.83, \textit{SD} = 6.22),  respectively.

A paired t-test was conducted to compare the total number of discoveries between sessions with Arti and Marti. The analysis revealed a significant difference in the number of discoveries measured, \textit{t}(11) = -3.25, \textit{p} = .0005**. A Mixed Repeated two-way ANOVA showed no significant order effect for this data.

\subsection{Delta Time of First Explicit Request}

In this section, we observed a notable difference in how users began their tasks between groups (P->NP) and (NP->P). Table \ref{delta} shows the time it took users to explicitly request for their first chart after the initial utterance.

\begin{table}[h]
\caption{This table illustrates how participants who started with Marti, were hesitant to explicitly request for their first chart, delaying their data exploration analysis.}
\label{delta}

\begin{center}
\begin{tabular}{cccc}
\toprule
 Session \#& Delta Time (m:s) &Session \#&Delta Time (m:s)\\
 \hline
     2   & 1:13    &1&   1:08\\
     4&   1:10  & 3   &0:57\\
     6 &3:29 & 5& 0:28\\
     8    &0:46 & 7&  0:24\\
     10&   2:15  & 9&0:59\\
     12& 1:44  & 11   &1:02\\
\bottomrule
\end{tabular}
\end{center}
\end{table}

While this may not seem like a significant finding at first, the discussion reveals its importance in Section \ref{ease}. We suspect that participants had a quicker and smoother experience getting accustomed to the system when they started with Arti. Those who interacted with Arti appeared to dive into their analysis immediately, whereas participants using Marti showed hesitation in getting started.

\subsection{Post-Interview}

\begin{figure}[ht]
  \centering
  \includegraphics[width=\columnwidth]{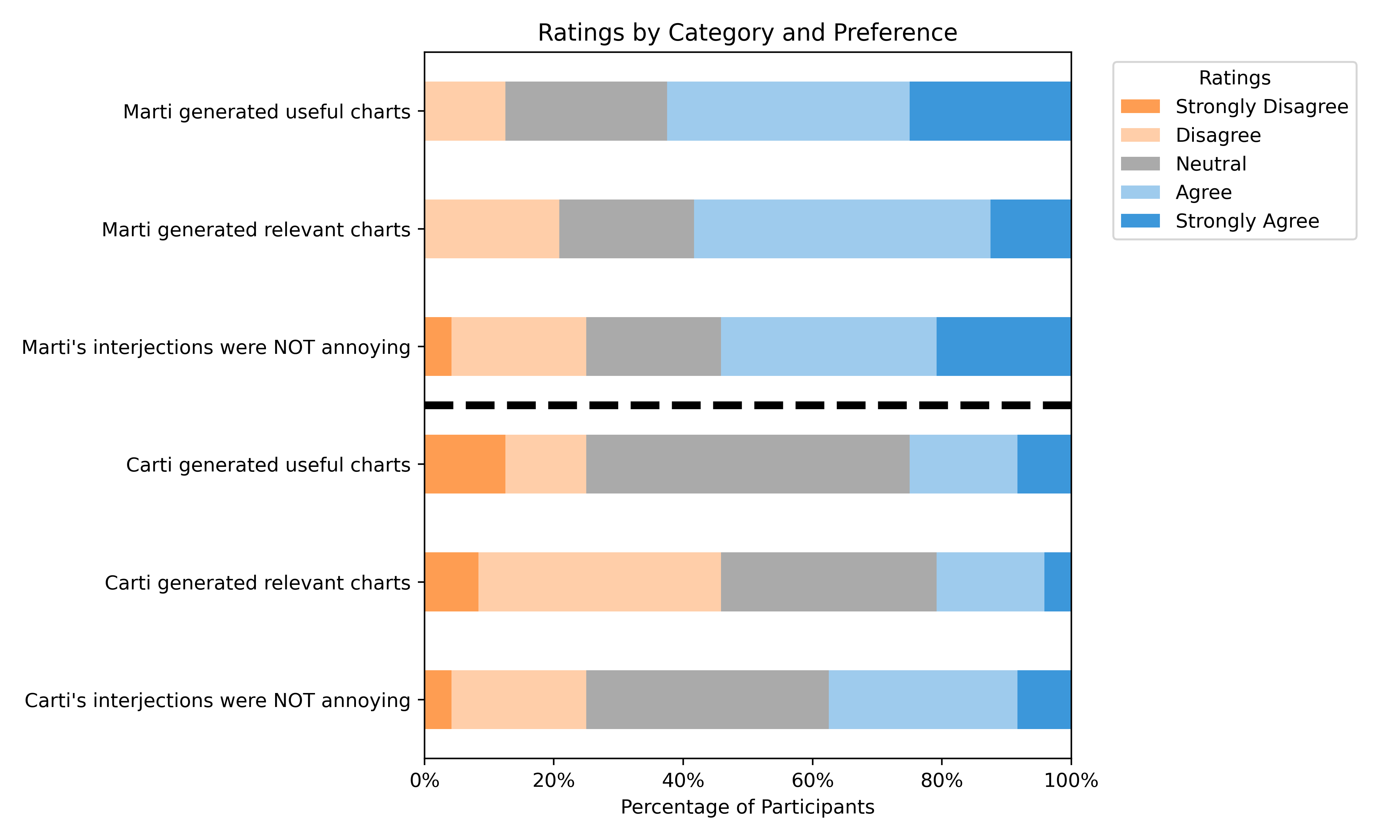}
  \caption{The top 3 bars represent the participants' feedback on Marti and the bottom 3 bars represent the participants' feedback on Arti. The figure indicates that users seemed to prefer their interaction with Marti.}
  \label{ratings}
\end{figure}

This section presents feedback gathered from participants during the post-interview, including their individual ratings and open responses from the semi-structured interviews. Figure \ref{ratings} shows the individual ratings from the participants where we asked the participants how they felt when they interacted with each digital assistant.

Here are the specific questions that we asked the participants to rate each assistant:
\begin{enumerate}
    \item On a scale from 1 to 5, with 1 being “extremely annoying” and 5 being “not annoying at all”, please rate your experience when the Arti/Marti interjected with charts
    \item On a scale from 1 to 5, with 1 being “irrelevant” and 5 being “very relevant”, please rate your experience when Arti/Marti interjected with a chart.
    \item On a scale from 1 to 5, with 1 being “not useful” and 5 being “very useful”, please rate your experience when Arti/Marti interjected with a chart. 
\end{enumerate}

\begin{figure}[ht]
  \centering
  \includegraphics[width=\columnwidth]{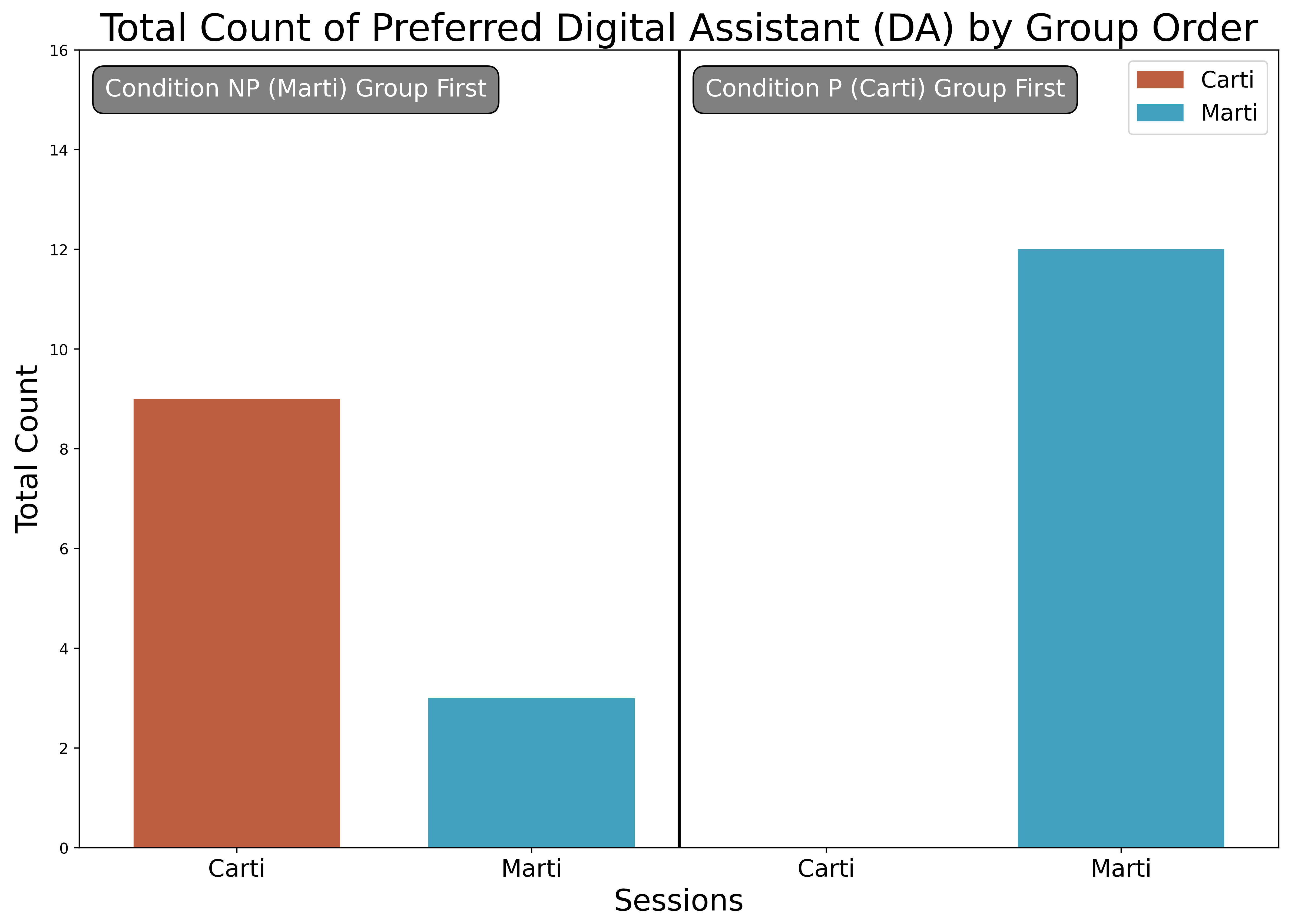}
  \caption{This chart displays the participants' digital assistant preference. The chart is separated between the two groups P (Arti) first and NP (Marti) first. Interestingly, all participants who interacted wtih Arti first, seemed to prefer Marti. And almost all of the participants preferred Arti when they interacted with Marti first.}
  \label{preferences}
\end{figure}

We asked the participants which digital assistant they preferred, Arti or Marti. Figure \ref{preferences} shows the participants' responses. 

We indicate the division into the two by-order groups (NP->P) and (P->NP) due to the significant order effects some of our data indicated. Out of 24 participants, 15 preferred Marti, while 9 favored Arti. Interestingly, 9 of the participants who preferred Arti initially interacted with Marti first.

\subsection{Qualitative Feedback}

Although most participants preferred Marti, those who liked Arti were predominantly from the group that used Arti second (NP->P). We also asked participants, "If you were to conduct a data exploration task, which digital assistant would you choose, and why?" The responses revealed a more nuanced preference, with several participants expressing interest in using both assistants. Below, we summarize the feedback into three categories:

\begin{description}
    \item[Participants who strictly chose Arti] These participants appreciated Arti’s proactive approach, noting that it provided charts to examine while they waited for requested ones. One participant highlighted that Arti allowed them to focus entirely on understanding the data without worrying about how to use the system. Another group mentioned that Arti felt faster and smarter than Marti, actively participating in the conversation rather than just passively listening.
    \item[Participants who strictly chose Marti] Some participants preferred Marti because they found Arti’s proactivity disruptive, breaking their train of thought. They felt that Arti was "too fast," sometimes generating relevant charts before they were ready or producing charts that seemed irrelevant.
    \item[Participants who chose both] A few participants expressed a desire to work with a "toned down" version of Arti. They believed there’s "a time for creativity and a time for focus." These participants appreciated Arti’s ability to generate ideas and help them get started but preferred Marti for more focused tasks once they were familiar with the system. One participant mentioned that working with a partner while interacting with Arti was challenging, as they had to consider their partner's input while also processing Arti’s proactive suggestions. They felt that working alone with Arti might have been more effective.
\end{description}

\section{Discussion}
Our results section highlighted the differences in the user interaction and outcomes when using a proactive digital assistant versus a non-proactive one during data exploration tasks (Research Question 1). In this section, we present our interpretation of the results in light of Research Question 2: "What are the benefits of interacting with a proactive digital assistant during data exploration?" We identified three key benefits when users engaged with Arti, the proactive assistant: improved system learnability and effectiveness, enhanced reliability of participants' findings, and greater variability in chart types used.

We close the section with an exploration of participants' acceptance of proactive agents and a discussion of potential future directions for the development of proactive digital assistants.

\subsection{Improved System Learnability and Effectiveness} \label{ease}

The results show that users made overwhelmingly more utterances when interacting with the proactive digital assistant, Arti. However, we saw a strong order effect: participants exposed to Marti, the non-proactive agent, first, and Arti second had more utterances on average towards Arti than Marti, but not significantly so, while the participants exposed to Arti first and Marti second had significantly more utterances towards Arti than Marti. This raises questions regarding the quality of the additional utterances and what could have caused the order effect: do participants used to Marti's behavior due to experiencing it first choose to use similar amount and style of verbiage, or do participants that had a chance to be exposed to Arti first felt less restricted during that session resulting in many utterances, and then later unconsciously limited their speech as Marti showed a more restrained behaviour?

As for the quality of utterances, further analysis of utterances that are relevant (use keywords) or that are connected to discoveries showed a significant improvement when talking to Arti in comparison to Marti, and this without an order effect. This indicates to us that the increase in utterances has a positive influence on task performance as there are more task-appropriate utterances.

Regarding the order effect, we suggest that exposure to Arti first allowed participants to be less restricted, while exposure to Arti second after Marti's restrained approach to chart generation, restricted those participants to some extent even during the session with Arti. Our observations suggest that participants who started with Arti, rather than Marti, found it easier to "get started" with their analysis. That means participants were quicker to decide which charts to request and to begin discussing the data. Arti facilitated this process by proactively generating charts for the participants that they could start with before any explicit requests. We highlight here that in sessions 6 and 10, participants who started with Marti took over 2 and 3 minutes, respectively, to generate their chart first. In contrast, the longest time it took participants who started with Arti was a minute and 8 seconds. This suggests that participants who started with Marti experienced a noticeable delay before feeling comfortable asking for charts, whereas those who started with Arti began their analysis more quickly. This delay persisted throughout the task, impacting how users continued to interact with the system and ultimately making it easier for them to learn how to use the system.Based on these results and observations, we believe that participants who started with the proactive assistant, Arti, had an easier time familiarizing themselves with the system and dataset, leading to higher overall efficiency.

Overall, in the context of speech-assisted data exploration tasks, proactive systems may have the potential to increase effectiveness (by encouraging unrestricted verbal interaction) and system learnability, a concept related to how users first interact with a system. This enables users to become quickly acquainted with using the system, allowing them to focus more on understanding the data and task at hand.

\subsection{Enhanced Reliability of Findings}

The proactive digital assistant, Arti, excelled at presenting the same data in different chart formats, which often helped users confirm or refute their hypotheses. For example, in session 12, participants were analyzing a box plot showing rainfall for Kauai. Initially, they quickly concluded that Kauai received no rainfall. However, Arti proactively generated a scatter plot showing rainfall and air temperature. Initially confused by data indicating instances of rainfall, the participants re-evaluated their conclusion and realized they had misinterpreted the original chart. This example demonstrates how a proactive system can help users validate or challenge their hypotheses, aligning with Sperber and Wilson's notion of "stimulating the cognitive environment" \cite{wilson1986defining}. In Pragmatics, stimulating the cognitive environment refers to how an utterance contributes meaningfully to the conversation. In contrast, when an utterance lacks usefulness, it may fail to spark interesting or productive thoughts.

This suggests that proactive systems have the potential to reassure users in the decisions and findings they make during data exploration. Similarly, Reicherts et al. developed "ProberBot," a chatbot designed to prompt users during investment decisions by engaging them in dialogue to justify their reasons for holding, buying, or selling stocks \cite{reicherts2022}. The study found that ProberBot encouraged users to take more time to re-evaluate their decisions, leading to more logical choices and reduced cognitive and emotional biases.

However, Reicherts et al. also observed that participants felt the bot’s dialogue could bias their decisions. In our case, while Arti can challenge users' hypotheses, we recognize the importance of ensuring that the assistant does not influence users toward incorrect beliefs. Like Reicherts et al., we emphasize that AI assistants must remain neutral in decision-making tasks and avoid steering users' thought processes.

\subsection{Greater Usage of Different Chart Types}

Arti broadened users' exploration of chart types. A common approach among users was to rely on a single chart type, with line charts being the most prevalent. In session 2, participants primarily used line charts with Marti in the first session, briefly switching to histograms near the end. However, when they interacted with Arti in the second session, Arti quickly generated a box plot—something the users had not explored with Marti. They continued using box plots throughout the session. A similar effect was observed in session 6, where participants initially used only line charts with Marti but explored scatter plots, box plots, histograms, and line charts with Arti, greatly expanding their chart usage. This variety helped participants draw more insights from the data.

In a previous study, Pins et al. examined how people interact with home voice assistants \cite{pins2020miss}. They found that systems like Google Assistant and Amazon Alexa are often underutilized, with users sticking to simple commands, especially after failed interactions. One issue is that these assistants' capabilities are either listed in a user manual or voiced upon request, making it difficult for users to remember them. In contrast, a visual display can remind users of a system’s capabilities. Our study shows the potential of proactive systems to suggest new features, helping users fully utilize the system’s capabilities.

\subsection{Users' Acceptance of a Proactive Assistant}

We found that users did not prefer to start with the proactive digital assistant. As illustrated in Figure \ref{preferences}, none of the participants who began with the proactive assistant preferred to continue working with it. According to feedback from post-interviews, starting with Arti may have been overwhelming for participants as they began the task. This is understandable given the multiple factors they had to manage: learning how to communicate with the digital assistant to create visualizations, coordinating with their partner on how to approach the climate data, and doing so in an unfamiliar environment. In contrast, participants who started with Marti had a chance to familiarize themselves with the system and dataset before transitioning to the proactive assistant, making the shift to Arti easier.

For those developing a proactive digital assistant for their own use cases, we advise caution in not overwhelming users. One participant noted that the proactive digital assistant was “too fast,” often creating useful charts before they even asked for them. While our findings suggest that starting with a proactive digital assistant might help users learn the system more quickly, it’s clear that users do not prefer to start with it. Future work could investigate ways for the assistant to adjust its level of proactivity based on the user's needs. Prior studies have explored "context-aware" systems, where computers measure the user's surroundings using technologies like computer vision, sensors, or temperature monitoring \cite{schmidt1999there, carvalho2017quality, ferrero2019ubiquitous}. For instance, one could imagine using a camera to observe the user's face, eyes, mouse movements, or body language to detect whether the user has absorbed the information before providing more.

However, in the case where users preferred to end with the proactive assistant, we observed there to be potential decline regarding whether the proactive assistant’s input is accepted. In the cases where the participants ended with the proactive agent, the users may have already found a workflow that works best for them. This was very apparent in sessions 4 and 8, where the participants started with the non-proactive assistant in task 1, and ended with the proactive assistant in task 2. The participants found a strategy for solving task 1. They continued to use this strategy in task 2, where they may have ignored more of what the agent was proactively generating. As opposed to sessions that started with the proactive assistant in task 1, where the users were more open to accepting what the assistant was generating.

\subsection{Future Directions for Proactive Assistants}


A future implementation that may improve the experience working with a proactive assistant may be to separate the content of what the proactive assistant generates and what users explicitly asked for. The participants in session 10 mentioned that it was difficult to differentiate between what the digital assistant generated and what they explicitly asked for. By separating what the assistant proactively generates and what the users explicitly ask for, the user can either focus on their data exploration without being interrupted by what the assistant generates. Then they are given the option to see what the assistant generates for further exploration. Andolina et al., question the possibility of combining proactive content and explicitly generated search results in the same interface \cite{andolina2018investigating}. Our study reveals that in the context of data exploration, having proactively generated charts and explicitly generated charts in the same interface may cause some confusion. 


\section{Conclusion}

In this study, we developed and evaluated a proactive data exploration assistant designed to generate relevant content and support users during data analysis tasks. Our approach was motivated by Pragmatics, which posits that effective utterances are those that are relevant to the conversation. Our findings support this theory, as users made more task-related utterances and discoveries with the proactive assistant compared to the non-proactive one. Additionally, our results demonstrate how a proactive assistant can improve system learnability, enhance the reliability of users' findings, and increase engagement with the system’s charting capabilities.
While the results indicated that users could potentially be more efficient with the proactive assistant, as it helped them initiate analysis more quickly through automatic chart generation, further investigation is needed. Specifically, a between-subject design would be more suitable for thoroughly assessing the potential efficiency benefits.
Interestingly, while users were more efficient with the proactive assistant, many expressed a preference for less proactivity. This suggests a delicate balance between offering assistance and allowing users to maintain control over their workflow. When the proactive assistant became too intrusive, users tended to ignore its input in favor of their own strategies. However, reducing proactivity too much could limit the assistant's ability to prompt new discoveries and guide users toward exploring more diverse chart types.
Our findings suggest that the proactive assistant positively impacted both task efficiency and the depth of user exploration. As technology continues to advance, there is an opportunity to refine natural language systems to more closely emulate human communication, as described by linguistic theories like Pragmatics. This could enable more intuitive and seamless interactions between users and digital systems.

\section{Acknowledgments}

This work was supported in part by National Science Foundation awards 2149133 and 2004014.
 
\begin{acks}

\end{acks}

\bibliographystyle{ACM-Reference-Format}
\bibliography{sample-base}

\appendix
\end{document}